\documentclass{article}
\usepackage{amssymb, amsmath}
\usepackage{nomencl}
\usepackage{grffile} 
\makeglossary
\usepackage[english]{babel}
\usepackage{setspace}
\printnomenclature 
\usepackage{color}
\usepackage{graphicx}
\usepackage{float}
\usepackage{amssymb}
\usepackage{amsmath}
\usepackage{epsfig}
\usepackage{subcaption} 
\usepackage[all]{xy}
\usepackage{psfrag}

\usepackage{amsthm,latexsym,multibox,amssymb}
\usepackage[utf8,latin1]{inputenc}
\inputencoding{utf8}
\usepackage{latexsym}
\makeatletter

\theoremstyle{plain}

\numberwithin{equation}{section} 
\numberwithin{figure}{section} 
\theoremstyle{plain}
\theoremstyle{definition}

\theoremstyle{definition}
\usepackage{verbatim}
\theoremstyle{plain}
 
\theoremstyle{plain}

\theoremstyle{remark}

\theoremstyle{remark}

\newcommand{\bea}{\begin{eqnarray}}
\newcommand{\eea}{\end{eqnarray}}

\usepackage{babel}

\makeatother
\makeglossary

\begin{document}
\title{The dynamics of polydisperse spray fuel \\in \\thermal explosion processes}

\author{\textbf{Shlomo Hareli}$^{a,*}$,\\ \textbf{OPhir Nave}$^{b}$\\    \textbf{Vladimir Gol'dshtein}$^{c}$\\ 
$^{a*}$ Department of Mathematics, Ben-Gurion University of the Negev,\\ Department of Mathematics, Azrieli College of Engineering 
\\$^{b}$ Department of Mathematics, College of Technology,
\\$^{c}$ Department of Mathematics, Ben-Gurion University of the Negev
\thanks{$E-mail\hspace{4pt}addresses$: harelish@post.bgu.ac.il (Sh. Hareli), Naveof@cs.bgu.ac.il (O. Nave), vladimir@bgu.ac.il (V. Gol'dshtein)}
}
\maketitle
\newpage
\begin{abstract} 
The dynamics of the particle-size distribution (PSD) of the polydispersed fuel is important for evaluation of the combustion process. In this paper we unfold the mystery of the dynamics represented by the PSD and gain new insight for the droplet behavior during the ignition process. A simplified model of the polydispersity is used for describing the system near the linear region, which evolves time depended function which describes the PSD. We examine the PSD fluctuation in time of the initial PSD described by the experimental as well as the Rosin-Rammler, Nukiyama-Tanasawa and the gamma distributions approximations. The results shows that the during the ignition process the droplets radii decrease as expected, yet the number of the smaller droplet increase in inverse proportion to the radius. An important novel result visualized by the graph, is the average radius of the droplets which first increases for shot time following by decreasing. The last result show the maximum of average radius is not located at the beginning of the process as expected. The algorithm presented here is superior to the parcel approach, since it can be applied for any approximation of the PSD, with negligible computation time.

\emph{Keywords:} {Polydisperse spray, Particle-size distribution dynamics, PDF Dynamics, Gamma distribution, Rosin-Rammler distribution, Nukiyama-Tanasawa distribution}
\end{abstract}

\newpage
\textbf{\Large{Nomenclature}}
\begin{singlespacing}
\begin{list}{}{}
\item $A$\hspace{20 pt}constant pre-exponential rate factor
\item $C$\hspace{20 pt}molar concentration ($k mol m^{-3}$)
\item $c$\hspace{20 pt}specific heat capacity ($Jkg^{-1}K^{-1})$)
\item $E$\hspace{20 pt}activation energy ($J k mol^{-1}$)
\item $L$\hspace{20 pt}liquid evaporation energy (i.e., latent heat of evaporation, Enthalpy of evaporation) ($Jkg^{-1}$)
\item $n_{d_{i}}$\hspace{20 pt}number of droplets of size $i$ per unit volume ($m^{-3}$)
\item $Q$\hspace{20 pt}combustion energy ($J kg^{-1}$)
\item $B$\hspace{20 pt}universal gas constant ($J k mol^{-1} K^{-1}$)
\item $R_{d_{i}}$\hspace{10 pt}radius of size $i$ drops  ($m$)
\item $R_{max, 0}$\hspace{20 pt}maximal droplet radius at $t=0$ ($m$)
\item $T$\hspace{20 pt}temperature ($K$)
\item $t$\hspace{20 pt}time ($s$)
\item $t_{react}$\hspace{20 pt}$A^{-1}e^{1/\beta}$ ($s$)
\item $P(\cdot)$ probability density function
\item $\hat{P}(\cdot)$ probability distribution function
\item $U(\cdot)$ Unit step function
\item $\Pi(\cdot)$ Rectangle function
\end{list}
\end{singlespacing}
\textbf{Greek symbols and dimensionless parameters}
\begin{singlespacing}
\begin{list}{}{}
\item $\alpha$\hspace{20 pt}dimensionless volumetric phase content
\item $\nu$\hspace{20 pt}the quantity equivalent to the volumetric phase content for the continuous model 
\item $\beta$\hspace{20 pt}dimensionless reduced initial temperature (with respect to the so-called activation temperature $E/B$)
\item $\gamma$\hspace{20 pt}dimensionless parameter that represents the reciprocal of the final dimensionless adiabatic temperature of the thermally insulated system after the explosion has been completed
\item $\epsilon_{1},\hspace{2 pt}\epsilon_{2} $\hspace{20 pt}dimensionless parameters introduced in Eq. \ref{eq:Parameters} and describing the interaction between gaseous and liquid phases 
\item $\mu$\hspace{20 pt}molar mass ($kg kmol^{-1}$)
\item $\lambda$\hspace{20 pt}thermal conductivity ($W m^{-1} K^{-1}$)
\item $\rho$\hspace{20 pt}density ($kg m^{-3}$)
\item $\tau$\hspace{20 pt}dimensionless time
\item $\Psi$\hspace{20 pt}represents the internal characteristics of the fuel (the ratio of the specific combustion energy and the latent heat of evaporation) and defined in equation \ref{eq:Parameters} (dimensionless)
\end{list}
\end{singlespacing}
\textbf{Dimensionless variables}
\begin{singlespacing}
\begin{list}{}{}
\item $\eta$\hspace{20 pt}dimensionless fuel concentration
\item $\theta$\hspace{20 pt}dimensionless temperature
\item $r$\hspace{20 pt}dimensionless radius
\end{list}
\end{singlespacing}
\textbf{Subscripts}
\begin{singlespacing}
\begin{list}{}{}
\item $d$\hspace{20 pt}liquid fuel droplets
\item $f$\hspace{20 pt}combustible gas component of the mixture
\item $g$\hspace{20 pt}gas mixture
\item $i$\hspace{20 pt}number of droplet sizes
\item $L$\hspace{20 pt}liquid phase
\item $p$\hspace{20 pt}under constant pressure
\item $s$\hspace{20 pt}saturation line (surface of droplets)
\item $0$\hspace{20 pt}initial state
\item $m$\hspace{20 pt}number of droplet sizes
\end{list}
\end{singlespacing}

\newpage
\section{introduction} 
Studying the dynamics of the particle-size distribution help to give a more comprehensive insight of the combustion process. Even that dynamics is an important aspect, there is very little work done on this subject \cite{Dynamics_Aggrawal_1998,Dynamics_Stone_1998,Dynamics_Sazhin_2000}, due to computation constraints. Most of the rich mechanism of the combustion is still unknown, knowing the dynamics allows more insight. 

For approximating of the experimental PSD, there are three well known methods commonly used, Rosin-Rammler distribution, Nukiyama-Tanasawa distribution and Gamma distribution \cite{Sum_2Integral_Modeling_Drop_Distributions_Babinsky_2002, Diffrent_PDF_Approx_RR_NT_G_Most_Used_Urban_2018}. These approximations provide comparatively accurate fitting for experimental data while allows extrapolation of for values beyond the measurement scope. These approximations permit easy calculation of parameter of interest since they are built-in function in software for calculation (e.g. MATLAB, MATHEMAICA) and with known associate integrals. We mention that there is no single mathematical expression that fit accurately the experimental data. Despite the fact that it has no theoretical basis, the Rosin-Rammler is the most widely used particle size distribution. There is also a desire to calculate the system dynamic based on the raw experimental data.

Most attempts to describe the combustion process has resulted in complex systems of partial differential equations\cite{simultaneously_processes_Warnatz_2006}. The main difficulty is to account for the many radii of the droplet that the composes the fuel. Semenov pioneered the description of thermal explosion theory by mathematical differential equation \cite{first_Math_Semenov_1928}.  Models that followed Semerov work that describe the combustion process that took in account the various droplet radii tend to be complex \cite{SMD_Over_Est_Ignition_Parcel_Bykov_2007}. The 'parcel' method divide the droplets into section but due to computation complexity this approach fail to estimate a realistic distribution . In \cite{Dep_PDF_Nave_2010} was purposed a model that described the polydespersed combustion model using a PDF which is a function of one variable the drop with the maximal radius. The last model allow us to monitor different PDF and their evolution with time.

\section{model description}

First we present the constraints on the combustion model. Secondly, we present the models which incorporates probability function which corresponds uniquely by the maximum radius. 
 
Several assumptions is used for acquiring model which describes the thermal explosion of vaporized fuel droplets. The system is taken to be adiabatic as the ignition process occurs briefly in comparison to heat loss by diffusion. The pressure variations showed to be negligible as to the mean pressure of the system \cite{first_Math_Semenov_1928}. In addition we assume that diffusion of the of the gas phase is inferior in comparison to the liquid phase, as a result in the heat transfer coefficient is determined by gas features \cite{Diffusion_Gas_Inferior_2_Liquid_Sazhin_2008}. We further assume the liquid temperature is identical to the temperature of the saturated liquid as the quasi-steady state approximation holds for the vaporizing droplets \cite{Quasi_SS_Approx_Spray_William_1985}. Finally the reraction is modeled as first order, highly exothermic chemical reaction leading to description by system of non-linear ordinary differential equations. Following the listed assumption the system of equations that describe the phenomena is given below \cite{SMD_Over_Est_Ignition_Parcel_Bykov_2007} .

\begin{equation}
\alpha_{g}\rho_{g}c_{pg}\frac{dT_{g}}{dt}=\alpha_{g}\mu_{f}Q_{f}AC_{f}e^{(-\frac{E}{BT_{g}})} - 4\pi\lambda_{g}(T_{g}-T_{d})\sum^{m}_{i=1}R_{d_{i}}n_{d_{i}},
\label{eq:Energy Equation polydisperse}
\end{equation} 

\begin{equation}
\frac{d(R^{2}_{di})}{dt}=\frac{2\lambda_{g}}{\rho_{L}L}(T_{d}-T_{g}),\hspace{4 pt}i=1,...,m,\hspace{4 pt}(m\hspace{4 pt}equations)\label{eq:Mass Equation polydisperse}
\end{equation} 

\begin{equation}
\frac{dC_{f}}{dt}=-AC_{f}e^{(-\frac{E}{BT_{g}})} + \frac{4\pi\lambda_{g}}{L\alpha_{g}\mu_{f}}(T_{g}-T_{d})\sum^{m}_{i=1}R_{d_{i}}n_{d_{i}}.
\label{eq:Concentration Equation polydisperse}
\end{equation}  

The initial conditions are:

\begin{equation}
at\hspace{4 pt}t=0:\hspace{4 pt}T_{g}(t=0)=T_{g0},\hspace{4 pt}R_{d_{i}}=R_{d_{i0}},\hspace{4 pt}\forall i: T_{d_{i}}=T_{g0},\hspace{4 pt} C_{f}=C_{f0}. 
\label{eq:Initial Conditions at time t polydisperse}
\end{equation}

For the model above the droplet radius distribution can be approximated by a continuous probability density function. The droplets radii sum is replaced by integral of the corresponding PDF \cite{Sum_2Integral_Modeling_Drop_Distributions_Babinsky_2002}. We use the sub index 0 to represent condition at time t=0.  

\begin{equation}
\int^{\infty}_{0}R_{0}\hat{P}_0(R_{0})dR_{0}=\sum_{i=1}^{m}n_{d_{i}}R_{d_{i,0}}.
\label{eq:PDFcontinuous}
\end{equation}

The were the normalized probability density function is given by:

\begin{equation}
P_{0}(R_{0})=\frac{\hat{P}_0(R_{0})}{\left\langle\hat{P}_0\right\rangle},\hspace{6 pt}were;\hspace{6 pt}\left\langle \hat{P}_0\right\rangle\equiv\int_{0}^{\infty}\hat{P}_0(R_{0})dR_{0}.
\label{eq:PDFcontinuousNormalization}
\end{equation}

Note that the integration by variable $R_{0}$ correspond to integration over the all the radius of the droplets at the initial time. The droplets estimation by PDF is superior to the parcel approach as the continuous PDF allows taking into account droplets not appearing on the experimental data. In addition the PDF is and not suffering by the limitation of the parcels method which describes the droplet distribution poorly since it can use small number of parcels due to calculation complexity.
\newpage

Using \eqref{eq:Mass Equation polydisperse} the radius R can be expressed as follows: 

\begin{equation}
R=\left(R_{m}^{2}-R_{m,0}^{2}+R_{0}^{2}\right)^{(1/2)}.
\label{eq:general radius explicit expression}
\end{equation} 

Thus $R_{0}$ can be obtained as a function of R.
\begin{equation}
R_{0}=G\left(R\right)\equiv\left(R^{2}-(R_{m}^{2}-R_{m,0}^{2}\right)^{(1/2)}.
\label{eq:general radius explicit expression for R0}
\end{equation} 

We define the rectangle function:
\begin{equation}
\Pi_{a}(x)\equiv{U_{0}(x)-U_{a}(x)}.
\label{eq:expression for rectangle function}
\end{equation} 

Using the last expressions we have:
\begin{eqnarray}
F(R_{m})\equiv\sum^{m}_{i=1}R_{d_{i}}n_{d_{i}}=\int^{R_{max, 0}}_{0}R_{0}P(R_{0})dR_{0}=\nonumber\\
\int^{\infty}_{0}G(R) P_{0}(G(R))\Pi_{R_{m}}(R)dR
\label{eq: final approximation}
\end{eqnarray}

Finally we can obtain the non dimensional model using Semenov's theory given by the parameters and transformation of variables: 

\begin{eqnarray}
\tau=\frac{t}{t_{react}},\hspace{4 pt}t_{react}=A^{-1}e^{\left(\frac{E}{BT_{g0}}\right)},\hspace{4 pt}\beta=\frac{BT_{g0}}{E},\hspace{4 pt}\gamma=\frac{c_{pg}T_{g0}\rho_{g0}}{C_{f0}Q_{f}\mu_{f}}\beta,\nonumber\\
\theta=\frac{E}{BT_{g0}}\frac{T_{g}-T_{g0}}{T_{g0}},\hspace{4 pt}\eta=\frac{C_{f}}{C_{f0}},\hspace{4 pt}r_{m}=\frac{R_{m}}{R_{m, 0}},\hspace{4 pt}\Psi=\frac{Q_{f}}{L}\label{eq:Parameters}\\ 
\epsilon_{2}=\frac{Q_{f}C_{f0}\alpha_{g}\mu_{f}}{\rho_{l}\nu_{0} L},\hspace{4 pt}\epsilon_{1}=\frac{4\pi\lambda_{go}R_{m, 0}\left\langle \hat{P}_0\right\rangle\beta T_{g0}}{AQ_{f}C_{f0}\alpha_{g}\mu_{f}}e^{\left(\frac{E}{BT_{g0}}\right)},\nonumber\\
\nonumber
\end{eqnarray}

\newpage

We shell rewrite $F(R_{m})$ as a function of $r_{m}$.\\
Notice: 

\begin{equation}
\begin{aligned}
G\left(R\right)=\left(R^{2}-(R_{m}^{2}-R_{m,0}^{2}\right)^{(1/2)}=\\R_{m, 0}\left\{\left(\frac{R}{R_{m,0}}\right)^2-\left(\left(\frac{R_{m}}{R_{m, 0}}\right)^2-1\right)\right\}^{1/2}=\\R_{m, 0}\left\{\left(\frac{R}{R_{m,0}}\right)^2-\left(r_{m}^2-1\right)\right\}^{1/2}
\end{aligned}
\label{eq:G in new coordinate}
\end{equation} 

Thus we have:
\begin{equation}
\label{eq:F_tilda}
\begin{gathered} 
R_{m, 0}\tilde{F}(r_{m}) \equiv F(R_{m})=\int^{\infty}_{0}G(R) P_{0}(G(R))\Pi_{R_{m}}(R)dR=\\R_{m, 0}\int^{\infty}_{0}\left\{\left(\frac{R}{R_{m,0}}\right)^2-\left(r_{m}^2-1\right)\right\}^{1/2}P_{0}\left(R_{m, 0}\left\{\left(\frac{R}{R_{m,0}}\right)^2-\left(r_{m}^2-1\right)\right\}^{1/2}\right)\Pi_{r_{m}\cdot R_{m,0}}(R)dR \\ \Rightarrow \tilde{F}(r_{m})=\frac{1}{R_{m,0}}F(r_{m}, R_{m,0}).
\end{gathered}
\end{equation}

The new equations are:
\begin{equation}
\gamma\left(1+\beta\theta\right)^{-1}\frac{d\theta}{d\tau}=\eta e^{(\frac{\theta}{1+\beta\theta})}-\epsilon_{1}\left(1+\beta\theta\right)^{1/2}\theta\tilde{F}(r_{m}),
\label{eq:Energy Equation non-dimension}\end{equation}
\begin{equation}
\frac{d(r_{m}^{2})}{d\tau}=-\frac{2}{3}\epsilon_{1}\epsilon_{2}\left(1+\beta\theta\right)^{1/2}\theta, 
\label{eq:Mass Equation nondimension}\end{equation} 
\begin{equation}
\frac{d\eta}{d\tau}=-\eta e^{(\frac{\theta}{1+\beta\theta})}+\epsilon_{1}\Psi\left(1+\beta\theta\right)^{1/2}\theta\tilde{F}(r_{m}),
\label{eq:Concentration Equation nondimension}
\end{equation}
where: $\tilde{F}(r_{m})$ is a function of the maximal dimensionless radius $r_{m}$ given in \eqref{eq:F_tilda}.\\
The non-dimensional initial conditions are:
\begin{equation}
at\hspace{4 pt}\tau=0:\hspace{4 pt}\theta=0,\hspace{4 pt}r_{m}=1,\hspace{4 pt}\eta=\eta_{0}. 
\label{eq:Initial Conditions}\end{equation}

\newpage
\section{Results and Discussion}
\begin{figure}[htbp]
	\centering
		\includegraphics[width=0.9\textwidth]{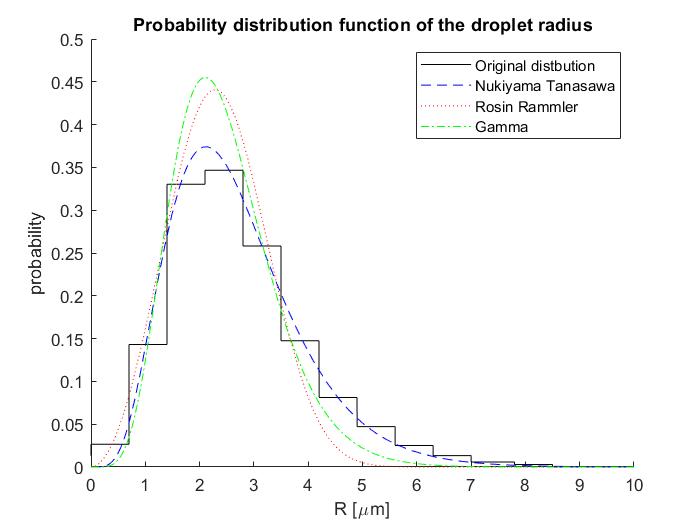}
	\caption{Normalized experimental probability and various approximation probability functions of the droplets radius}
	\label{fig:PDF}
\end{figure}

Three widely used distribution for describing the experimental droplets radii are Rosin-Rammler distribution \cite{PDF_Valid_Rosin_1933}, Nukiyama-Tanasawa distribution \cite{Nukiyama_Tanasawa_1939} and the well known Gamma distribution. These distribution parameters was chosen to fit the experimental data. The experimental distribution and the distributions that were used to approximate the experimental distribution are normalized and presented below (figure ~\ref{fig:PDF}). 

\begin{figure*}
        \centering
        \begin{subfigure}[b]{0.475\textwidth}
            \centering
            \includegraphics[width=\textwidth]{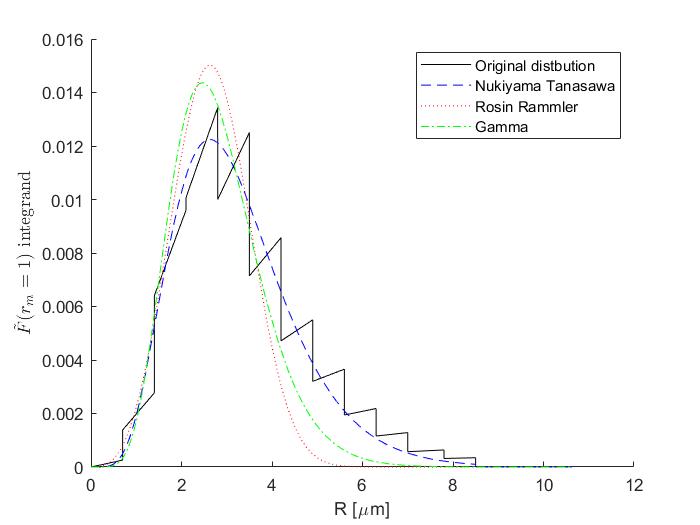}
            \caption[Network2]%
            {\small The integrand of $\tilde{F}$ at $r_{m} =1$}  
            \label{fig:F-tilda at rm=1}
        \end{subfigure}
        \hfill
        \begin{subfigure}[b]{0.475\textwidth}  
            \centering             \includegraphics[width=\textwidth]{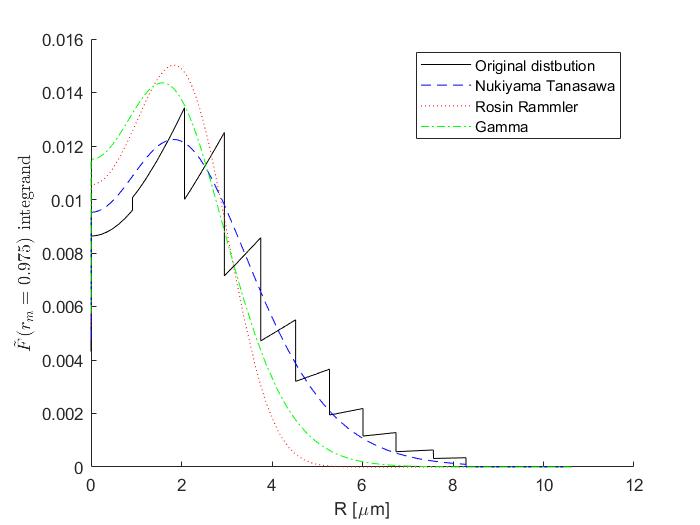}
            \caption[]%
            {\small The integrand of $\tilde{F}$ at $r_{m} =0.975$}   
            \label{fig:F-tilda at rm=0.975}
        \end{subfigure}
        \vskip\baselineskip
        \begin{subfigure}[b]{0.475\textwidth}   
            \centering 
            \includegraphics[width=\textwidth]{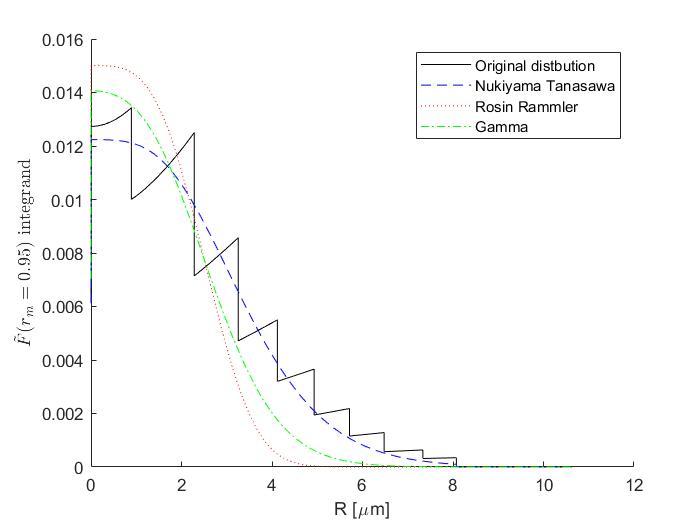}
            \caption[]%
            {\small The integrand of $\tilde{F}$ at $r_{m} =0.95$}   
            \label{fig:F-tilda at rm=0.95}
        \end{subfigure}
        \quad
        \begin{subfigure}[b]{0.475\textwidth}   
            \centering 
            \includegraphics[width=\textwidth]{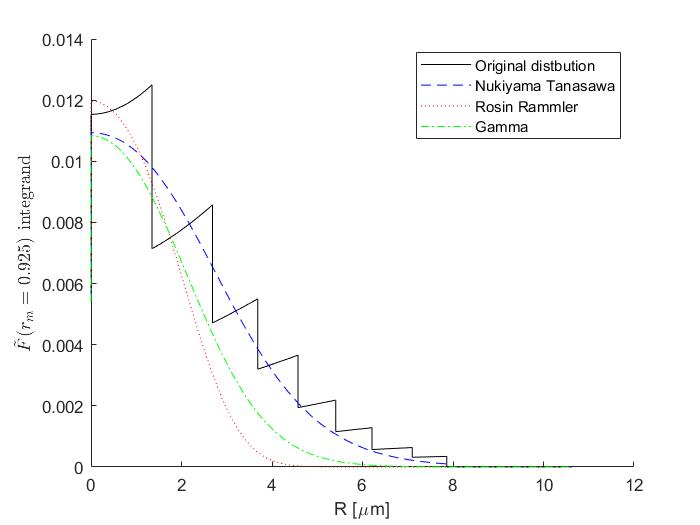}
            \caption[]%
           {\small The integrand of $\tilde{F}$ at $r_{m} =0.925$}   
            \label{fig:F-tilda at rm=0.925}
        \end{subfigure}
        \caption[ The average and standard deviation of critical parameters ]
        {\small The integrand of $\tilde{F}$ at region near $r_{m}=1$ for experimental data and its approximations, at various values of $r_{m}$} 
        \label{fig:F-tilda near rm=1}
    \end{figure*}

Near $r_{m} =1$ there is a semi linear behavior of $r_{m}$ as a function of time. Close to the linear region our model is valid, thus we can study the various functions dynamic, which describes particle-size distribution. The integrand of F represent the PSD, this clearly given by definition of F (~\ref{eq: final approximation}), While the integrand of $\tilde{F}$ is proportional to the PSD (as can be seen by equation ~\ref{eq:F_tilda}). The dynamic of the PSD based on the initial measured PSD and it's approximation for various    
As can be seen (figure ~\ref{fig:F-tilda near rm=1} ) the  behavior of integrand of $\tilde{F}$ is similar.The experimental PSD is discontinuous as it discrete, while the approximation which are continuos .The integrand of F consist of $P_{0}(G(R))\Pi_{R_{m}}(R)$ that can be identified with the probability function (i.e. proportional to the probability function) . The change dynamic of the probability function is given below (figure ~\ref{fig:probability for various $r_{m}$ values}).

\begin{figure}[htbp]
	\centering
		\includegraphics[width=0.9\textwidth]{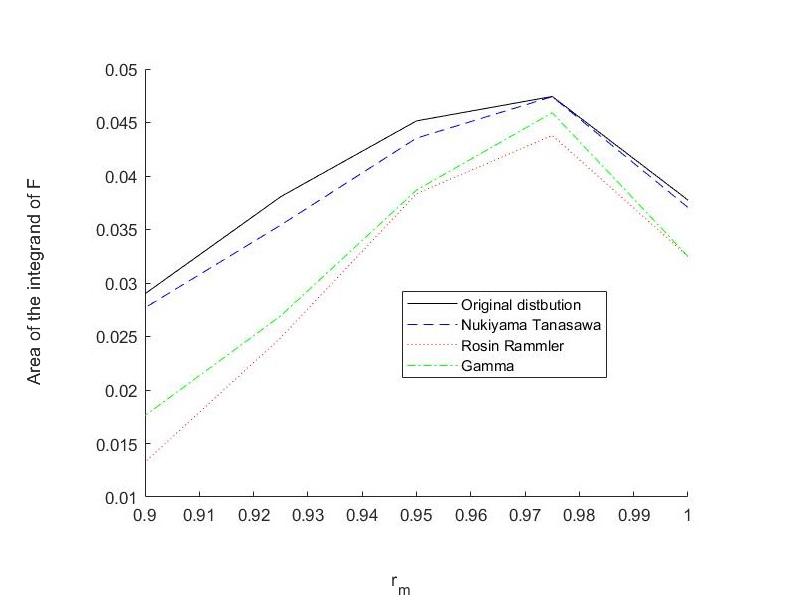}
	\caption{Integrand of F}
	\label{fig:F}
\end{figure}

\begin{figure}[htbp]
	\centering
		\includegraphics[width=0.9\textwidth]{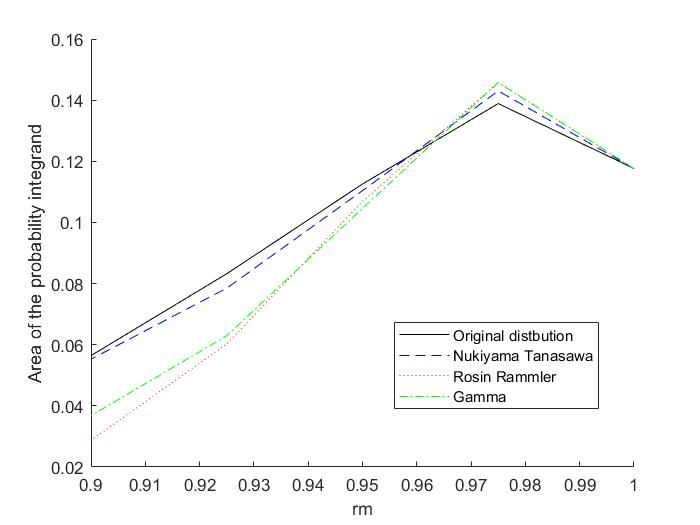}
	\caption{Probability}
	\label{fig:P}
\end{figure}

Figure ~\ref{fig:F} represent the area of the integrand of F that was originated by the initial PSD as a function of various values of $r_{m}$. The area is the F function value which corresponds to the droplet mean radius. As the maximum radius decreases it accepted that the mean radius will decrease as well. We notice initially an increase of the mean radius for a short period following by the expected behavior (i.e. the decrease of the mean radius). A similar behavior can be noticed also at the area of the changing probability function (figure ~\ref{fig:P}).This phenomena can be seen numerically, and hopefully will be derived analytically in the future.

\begin{figure*}
        \centering
        \begin{subfigure}[b]{0.475\textwidth}
            \centering
            \includegraphics[width=\textwidth]{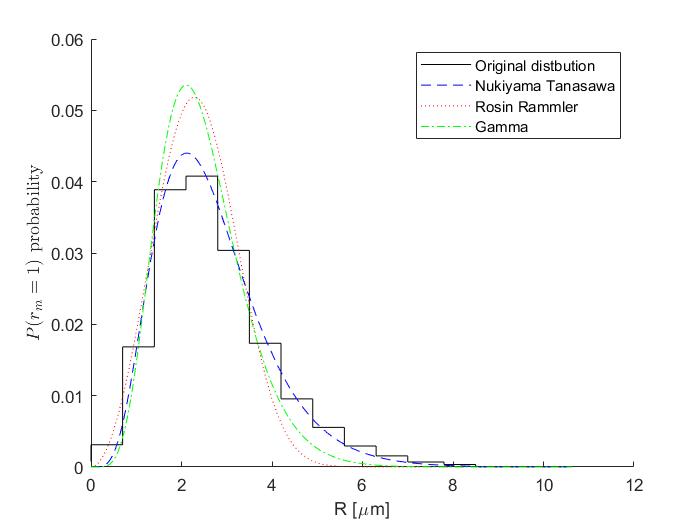}
            \caption[Network2]%
            {\small Probability at $r_{m}=1$}    
            \label{fig:Probability at $r_{m}=1$}
        \end{subfigure}
        \hfill
        \begin{subfigure}[b]{0.475\textwidth}  
            \centering 
            \includegraphics[width=\textwidth]{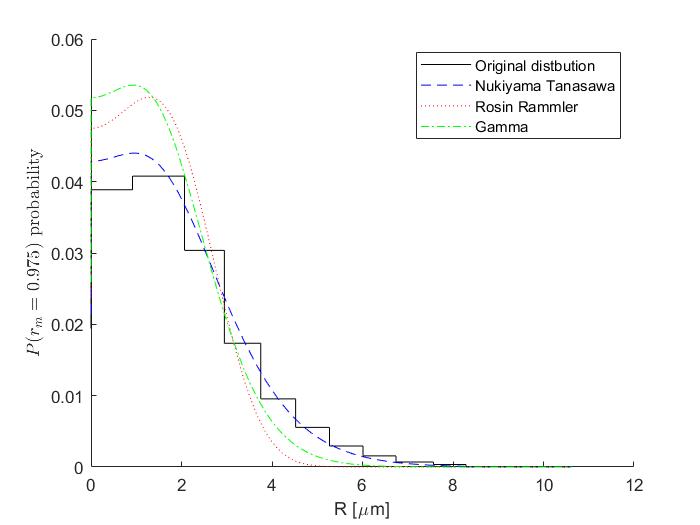}
            \caption[]%
            {\small Probability at $r_{m}=0.975$}    
            \label{fig:Probability at $r_{m}=0.975$}
        \end{subfigure}
        \vskip\baselineskip
        \begin{subfigure}[b]{0.475\textwidth}   
            \centering 
            \includegraphics[width=\textwidth]{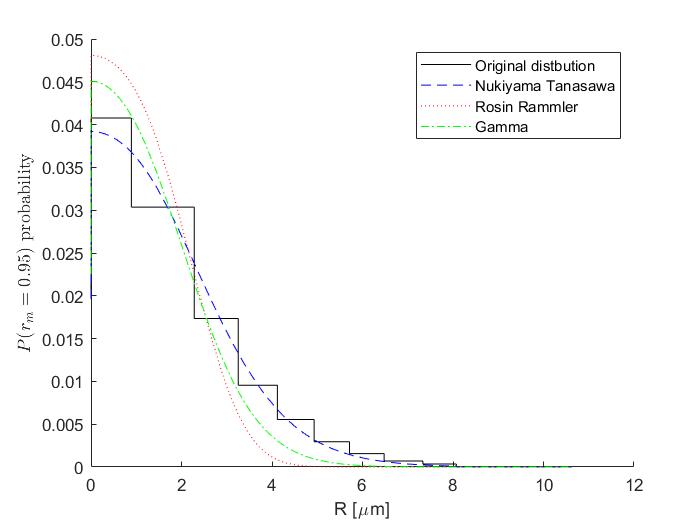}
            \caption[]%
            {{\small Probability at $r_{m}=0.95$}}    
            \label{fig:Probability at $r_{m}=0.95$}
        \end{subfigure}
        \quad
        \begin{subfigure}[b]{0.475\textwidth}   
            \centering 
            \includegraphics[width=\textwidth]{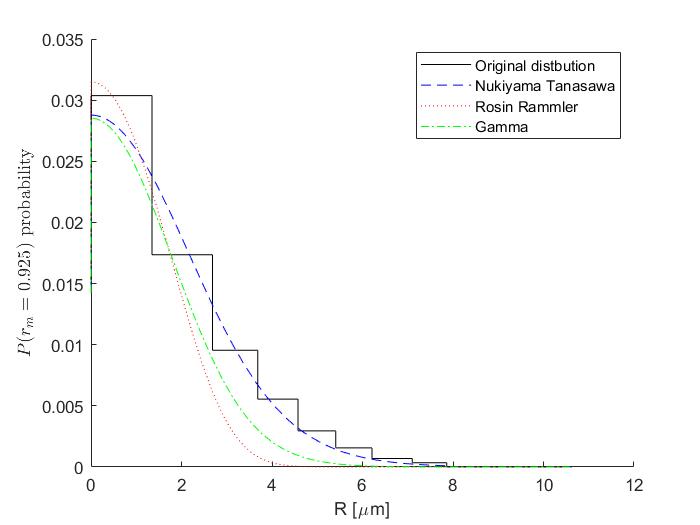}
            \caption[]%
            {{\small Probability at $r_{m}=0.925$}}    
            \label{fig:Probability at $r_{m}=0.925$}
        \end{subfigure}
        \caption[Probability at region near $r_{m} = 1$ for experimental data and its approximations, at various values of $r_{m}$]
        {\small Nukiyama-Tanasawa, Rosin-Rammler and Gamma probability for various $r_{m}$ values} 
        \label{fig:probability for various $r_{m}$ values}
    \end{figure*}

\begin{figure}[htbp]
	\centering
		\includegraphics[width=0.9\textwidth]{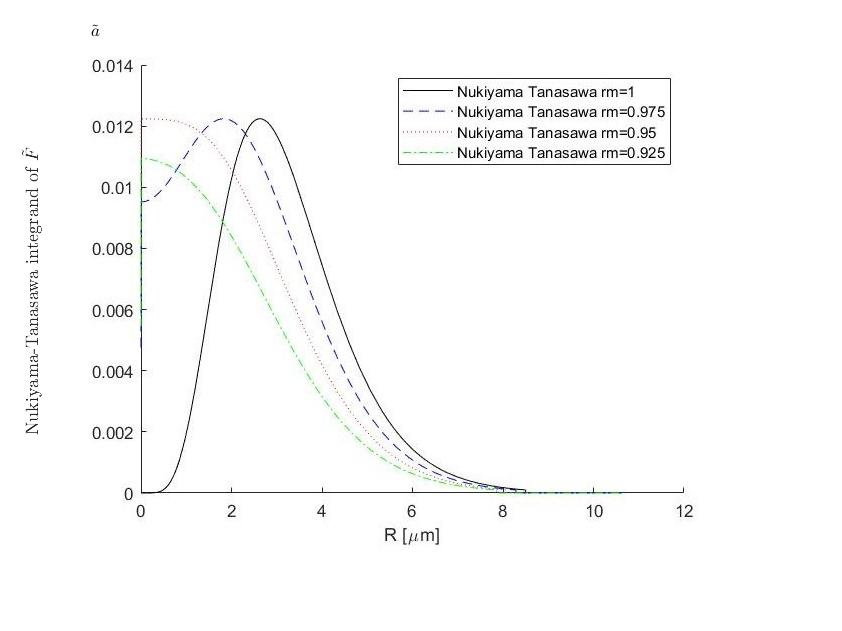}
	\caption{Nukiyama-Tanasawa integrand of $\tilde{F}$}
	\label{fig:NTF}
\end{figure}

Recal that integrand of F corresponds to the droplets radius distribution. For simplicity, we look at the integrand of F of the Nukiyama-Tanasawa approximation (figure~\ref{fig:NTF}). 
 The area of the probability integrand is understood if we concentrate on the Nukiyama-Tanasawa approximation. Let us rewrite equation \ref{eq:Mass Equation polydisperse} using $\frac{d(R^{2}_{di})}{dt}=2R_{di}\frac{d(R_{di})}{dt}$:
\begin{equation}
\frac{d(R_{di})}{dt}=\frac{1}{R_{di}}\frac{\lambda_{g}}{\rho_{L}L}(T_{d}-T_{g}),\hspace{4 pt}i=1,...,m,\hspace{4 pt}(m\hspace{4 pt}equations)\label{eq:new Mass Equation polydisperse}\end{equation} 
As $T_{d}<T_{g}$ since gas temperature is superior to the droplets temperature, the derivative is negative. The reduction of droplets radii in time is also valid since the radius is decreased in the presence of high temperature. As can be seen the curve of the droplets density is shifted to the right as time progresses as expected due to the reduction in size of the droplets. Another important observation is that each droplet reduction in radius is proportional to the inverse of it's radius. Last insight show that droplets of big redius will decrease in size more slowly than smaller radius droplets. This phenomena can be seen in the graph, where the numbers of small radius droplets rise as time increases. The phenomena can be explained by noticing that droplets with high surface area to volume ratio are more influenced by heat. Last observation shows that small droplets reduce size faster than larger droplets. Thus as the process progresses frequency smaller droplets will be increase as inverse proportion of radius.    
 
\newpage
\section{Conclusion}
The novel approach which was presented in this paper, helps to follow the dynamics of the PSD. Low computation time provide the ability to follow the dynamics of the experimental measured PSD as well as it's continuous approximations. The continuous approximation can allow exploring the behavior of the PSD analytically. The method show is superior to the parcel approach, which due to computation is limited to small amount of parcel. Our results show that the mean radius of the droplets eventually decrease but surprisingly initially increasing. We associate the phenomena of the increase of the average radius to tendency of smaller droplets to evaporate in higher rate. The ability to monitor the evolution in time of the complex system of droplets in the ignition process, open a window to new field in the combustion research. 
\newpage
\bibliographystyle{IEEEtran}
\bibliography{LaTex1AA}

\begin{thebibliography}{10}
\providecommand{\url}[1]{#1}
\csname url@samestyle\endcsname
\providecommand{\newblock}{\relax}
\providecommand{\bibinfo}[2]{#2}
\providecommand{\BIBentrySTDinterwordspacing}{\spaceskip=0pt\relax}
\providecommand{\BIBentryALTinterwordstretchfactor}{4}
\providecommand{\BIBentryALTinterwordspacing}{\spaceskip=\fontdimen2\font plus
\BIBentryALTinterwordstretchfactor\fontdimen3\font minus
  \fontdimen4\font\relax}
\providecommand{\BIBforeignlanguage}[2]{{%
\expandafter\ifx\csname l@#1\endcsname\relax
\typeout{** WARNING: IEEEtran.bst: No hyphenation pattern has been}%
\typeout{** loaded for the language `#1'. Using the pattern for}%
\typeout{** the default language instead.}%
\else
\language=\csname l@#1\endcsname
\fi
#2}}
\providecommand{\BIBdecl}{\relax}
\BIBdecl

\bibitem{Dynamics_Aggrawal_1998}
S.~K. Aggrawal, ``A review of spray ignition phenomena: present status and
  future research,'' \emph{Progress in Energy and Combustion Science}, vol.~24,
  no.~6, pp. 565--600, 1998.

\bibitem{Dynamics_Stone_1998}
R.~Stone, ``Introduction to {I}nternal {C}ombustion {E}ngines,''
  \emph{Macmillan, London}, 1998.

\bibitem{Dynamics_Sazhin_2000}
E.~M. Sazhin, S.~S. Sazhin, M.~R. Heikal, V.~I. Babushok, and R.~A. Johns, ``A
  detailed modeling of the spray ignition process in diesel engines,''
  \emph{Combustion Science 429 and Technology}, vol. 160, no.~1, pp. 317--344,
  2000.

\bibitem{Sum_2Integral_Modeling_Drop_Distributions_Babinsky_2002}
E.~Babinsky and P.~E. Sojka, ``Modeling drop size distributions,''
  \emph{Progress in Energy and Combustion Science.}, vol.~28, no.~4, pp.
  303--329, 2002.

\bibitem{Diffrent_PDF_Approx_RR_NT_G_Most_Used_Urban_2018}
A.~Urbán and V.~Józsa, ``Investigation of {F}uel {A}tomization with {D}ensity
  {F}unctions,'' \emph{Periodica Polytechnica Mechanical Engineering}, vol.~62,
  no.~1, pp. 33--41, 2018.

\bibitem{simultaneously_processes_Warnatz_2006}
J.~Warnatz, U.~Maas, and W.~R. Dibble, ``Combustion: {P}hysical and {C}hemical
  {F}undamentals, {M}odeling and {S}imulation, {E}xperiments, {P}ollutant
  {F}ormation,'' \emph{Springer-Verlag, Berlin Heidelberg, New York}, 2006.

\bibitem{first_Math_Semenov_1928}
N.~N. Semenov, ``Zur theorie des verbrennugsprozesse,'' \emph{Zeitschrift fur
  Physik}, vol.~48, pp. 571--581, 1928.

\bibitem{SMD_Over_Est_Ignition_Parcel_Bykov_2007}
V.~Bykov, I.~Goldfarb, and V.~M. G. B.~J. Greenberg, ``Auto-ignition of a
  polydisperse fuel spray,,'' \emph{Proceedings of the Combustion Institute},
  vol.~31, no.~2, pp. 2257--2264, 2007.

\bibitem{Dep_PDF_Nave_2010}
O.~Nave, V.~M. Gol’dshtein, and V.~Bykov, ``A probabilistic model of thermal
  explosion in polydisperse fuel spray,'' \emph{Applied Mathematics and
  Computation,}, vol. 217, no.~6, pp. 2698--2709, 2010.

\bibitem{Diffusion_Gas_Inferior_2_Liquid_Sazhin_2008}
S.~Sazhin, S.~Martynov, T.~Kristyadi, C.~Crua, and M.~R. Heikal, ``Diesel fuel
  spray penetration, heating, evaporation and ignition:{M}odelling vs.
  experimentation,'' \emph{International Journal of Engineering Systems
  Modelling and Simulation}, 2008.

\bibitem{Quasi_SS_Approx_Spray_William_1985}
F.~A. William, \emph{The {F}undamental {T}heory of {C}hemically {R}eacting
  {F}low {S}ystem}.\hskip 1em plus 0.5em minus 0.4em\relax Benjamin-Cummings,
  CA Menlo Park, 1985.

\bibitem{PDF_Valid_Rosin_1933}
P.~Rosin and E.~Rammler, ``Laws governing the fineness of powdered coal,.''
  \emph{Journal of the Institute of Fuel}, vol.~7, pp. 29--36, 1933.

\bibitem{Nukiyama_Tanasawa_1939}
S.~Nukiyama and Y.~Tanasawa, ``Experiments on the atomization of liquids in an
  airstream,'' \emph{Transactions of the Japan Society of Mechanical
  Engineers}, vol.~5, pp. 68--75, 1939.

\end{thebibliography}

\end{document}